\documentclass[conference]{IEEEtran}
\IEEEoverridecommandlockouts
\usepackage{cite}
\usepackage{amsmath,amssymb,amsfonts}
\usepackage{algorithmic}
\usepackage{graphicx}
\usepackage{textcomp}
\usepackage{xcolor}
\usepackage{booktabs}
\usepackage{caption}
\usepackage[T1]{fontenc}
\usepackage{lmodern} 
\usepackage{listings}
\usepackage{xspace}
\usepackage{url}
\usepackage{float}
\usepackage{subcaption}

\usepackage{multirow, multicol}
\def\BibTeX{{\rm B\kern-.05em{\sc i\kern-.025em b}\kern-.08em
    T\kern-.1667em\lower.7ex\hbox{E}\kern-.125emX}}
    
\definecolor{eclipse}{RGB}{127,0,85}

\newcommand{\sleec}[1]{{\small\fontfamily{qcr}\selectfont#1}}

\newcommand{\sleeckeyword}[1]{\textcolor{eclipse}{\textbf{#1}}}

\newcommand{\boldparagraph}[1]{\vskip 0.05in\noindent\textbf{#1.}}

\begin{document}

\title{Tool for Supporting Debugging and Understanding of  Normative Requirements Using LLMs}

\author{
\IEEEauthorblockN{
Alex Kleijwegt, 
Sinem Getir Yaman, 
Radu Calinescu}
\IEEEauthorblockA{Department of Computer Science, University of York, United Kingdom}
}


\maketitle

\section{Introduction}
Normative requirements specify social, legal, ethical, empathetic, and cultural (SLEEC) norms that must be observed by a system. To support the identification of SLEEC requirements, numerous standards and regulations have been developed (e.g.,\cite{eu_ai_act_2024a}). These requirements are typically defined by stakeholders in the non-technical system with diverse expertise (e.g., ethicists, lawyers, social scientists). Hence, ensuring their consistency and managing the requirement elicitation process are complex and error-prone tasks\cite{TOWNSEND2025PGYNC,troquard2024social}. Recent research has addressed this challenge using domain-specific languages to specify normative requirements as rules, whose consistency can then be analyzed with formal methods~\cite{Getir-Yaman-et-al-23,GETIRYAMAN2025RCCPT,feng-et-al-24-e,feng-et-al-24bb}.

Nevertheless, these approaches often present the results from formal verification tools in a way that is inaccessible to non-technical users. This hinders understanding and makes the iterative process of eliciting and validating these requirements inefficient in terms of both time and effort. 

To address this problem, we introduce SLEEC-LLM, a tool that uses large language models (LLMs) to provide natural-language interpretations for model-checking counterexamples corresponding to SLEEC rule inconsistencies~\cite{GETIRYAMAN2025RCCPT,Getir-Yaman-et-al-23}.
SLEEC-LLM improves the efficiency and explainability of normative requirements elicitation and consistency analysis. 
To demonstrate its effectiveness, we summarise its use in two real-world case studies involving non-technical stakeholders.

\section{Problem Description}

Our tool operates on SLEEC rules specified in SLEEC DSL, a domain-specific language for normative requirements~\cite{Getir-Yaman-et-al-23} 
that has proven accessible to stakeholders without a technical background (e.g., lawyers, ethicists, and regulators)~\cite{GETIRYAMAN2025RCCPT}.
The specification of a SLEEC rule set in SLEEC DSL comprises two parts: definitions and rules.

Definitions declare events and measures corresponding to system capabilities, and to activities during system interactions with the environment, including humans. 
\sleeckeyword{Events} correspond to instantaneous actions, whereas \sleeckeyword{measures} represent capabilities to provide (immediately) information captured by values of data types, such as \sleeckeyword{Boolean}, \sleeckeyword{numeric}, and \sleeckeyword{scale}. 
Rules have the basic form ``\sleec{\sleeckeyword{when} trigger \sleeckeyword{then} response}''. Such a rule defines the required response when the event in the trigger happens, and its conditions on measures, if any, are satisfied. 
A rule can be accompanied by one or more \emph{defeaters}, introduced using the ``\sleec{\sleeckeyword{unless}}'' construct. The language incorporates time constructs that allow responses with deadlines and timeouts using the ``\sleec{\sleeckeyword{within}}'' construct.

Given such a SLEEC rule set, an existing tool called SLEECVAL~\cite{Getir-Yaman-et-al-23} translates these rules into a formal language called \textit{tock CSP}, a time-variant process algebra, as illustrated in Fig.~\ref{fig:sleec-llm-workflow}. 
The formal semantics defined in CSP are then analyzed for consistency using the FDR model checker~\cite{FDR}. However, the results (particularly counterexamples) are difficult to interpret for non-technical stakeholders, e.g.,  
the rules 
``\sleec{R1} \sleec{\sleeckeyword{when} DetectUserFallen \sleeckeyword{then} CallEmergencySupport \sleeckeyword{within} 2 \sleeckeyword{minutes} \sleeckeyword{unless} emergencyLevel>L4 \sleeckeyword{then} CallEmergencySupport}'' and 
``\sleec{R2} \sleec{\sleeckeyword{when} DetectUserFallen \sleeckeyword{and} emergencyLevel<L2 \sleeckeyword{then} \sleeckeyword{not} CallEmergencySupport \sleeckeyword{within} 2 \sleeckeyword{minutes}}'' 
are conflicting, and the tool reports a counterexample in the following form where the \textit{tock} is a time unit.

\begin{lstlisting}[basicstyle=\small\ttfamily]
<DetectUserFallen, emergencyLevel.E1,
emergencyLevel.E1, tock, ..., tock>
\end{lstlisting}

\noindent
Such output is not easily understandable for non-technical users and lacks interpretation of exploratory resolution scenarios for the system, e.g., what type of error is occurring. Our LLM-enabled debugging and resolution tool (publicly available on GitHub~\cite{SLEEC-LLM}) addresses this problem.

\section{Tool Architecture and Workflow}
\boldparagraph{Prompt engineering and LLMs}
\begin{figure*}[htbp]
    \centering
\includegraphics[width=0.8\textwidth]{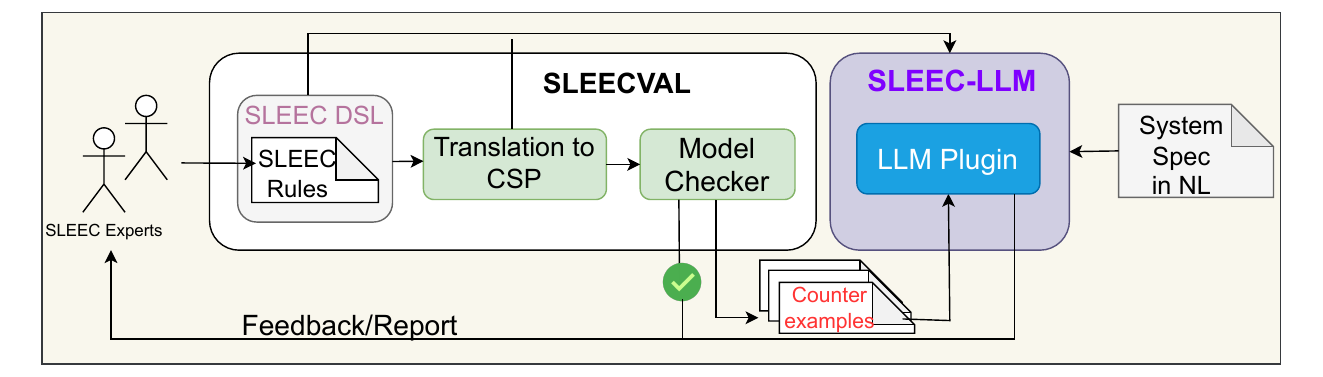}
    \caption{SLEEC-LLM workflow.}
    \label{fig:sleec-llm-workflow}
    \vspace*{-5mm}
\end{figure*}
Developing an effective prompt for an LLM is an iterative and flexible process that significantly impacts the quality of the LLM's output. The prompt evolves continuously, adapting to changes in input parameters like the number of counterexamples, available context, and the specific LLM model used.

The prompt begins by providing the LLM with comprehensive context including (Fig.~\ref{fig:sleec-llm-workflow}): (1)~SLEEC rules specified in SLEEC DSL for understanding outputs through RAG (Retrieval Augmented Generation), (2)~formal semantics in CSP, (3)~counterexamples, and (4)~a detailed natural-language description of the system. 

Following this context, the prompt includes a JSON file for the expected output, which is presented in Listing~\ref{lst:output}. The output is expected to include the errors and two different suggestions for resolution.

\section{Preliminary Results}
A user study was conducted to evaluate whether the SLEEC LLM tool helps resolve ruleset conflicts more quickly and with fewer development iterations. Using a control study among SLEEC participants (experts in law, ethics and psychology) who completed two tasks for two case studies.
The first group of experts used the SLEEC inconsistency resolution tool with the LLM for the ALMI case study, which involves a healthcare robot. The second group performed the same task for ASPEN, a forest-monitoring drone, also using the LLM tool.

\lstset{
  basicstyle=\footnotesize\ttfamily,
  columns=fullflexible,
  breaklines=true,
  backgroundcolor=\color{gray!10},
  frame=single
}

\begin{lstlisting}[caption={Output template for conflicting rules (partial)},label={lst:output}]
"Conflicting Rules":{
  "Error": {
    "Rule1": {name of Rule1},
    "Rule2": {name of Rule2},
    "Scenario":{a scenario that creates conflict in the rules},
    "Category": {deadlock, divergence, naming}
    "Justification": {your justification}
  }
   "Resolution": {add rule, combine rule, remove rule, modify rule
    "Suggestion1": {name of Rule1 A},
    "Justification": {your justification}
    "Suggestion2": {name of Rule2},
    "Justification": {your justification}
  }
} 
\end{lstlisting}

After completing their initial tasks, the groups switched case studies and repeated the tasks without the LLM tool. The tasks required participants to modify intentionally inconsistencies in SLEEC rulesets. Participants used the SLEEC tool to identify and fix rule conflicts iteratively, with a 25-minute time limit per task. When the LLM was used, participants could submit counterexample traces for suggestions.

Quantitative metrics included: 1) time to resolve conflicts (from counterexample to rule fix), 
2) number of iterations (SLEEC submissions until the ruleset was conflict-free)
The results are summarised in Table~\ref{tab:userstudy}.

\begin{table}[t]
\centering
\begin{tabular}{@{}p{3.7cm}p{1.8cm}p{1.1cm}p{1.1cm}@{}}
\toprule
\textbf{Task} & \textbf{Elapsed Time} & \textbf{Resolved \newline Rules} & \textbf{Iterations} \\ \midrule
Study 1: ALMI – Manual & Time expired & 2 / 4 & 6 \\
Study 2: ASPEN – Manual & Time expired & 2 / 4 & 8 \\
Study 1: ALMI – LLM assisted & 12m 30s & 4 / 4 & 4 \\
Study 2: ASPEN – LLM assisted & 25 m & 4 / 4 & 6 \\
\bottomrule
\end{tabular}
\caption{Results of user study on rule conflict resolution (4-rule set)}
\label{tab:userstudy}
\end{table}

\section{Discussion and Related Work}
We present a tool to support SLEEC experts in debugging and understanding inconsistencies in normative requirements during the requirement validation process. Existing research has focused on consistency validation methods~\cite{GETIRYAMAN2025RCCPT,feng-et-al-24-e}, and large language models (LLMs) have been used to understand the semantic relationships within SLEEC capabilities to produce more accurate results in the analysis~\cite{feng-et-al-24bb}. We believe our tool represents a promising direction for increasing the explainability and understandability of requirement elicitation inconsistencies for non-technical users. In the future, we plan to extend our work to support the verification of systems against SLEEC requirements.
\bibliographystyle{IEEEtran}
\bibliography{refs}

\begin{thebibliography}{1}
\providecommand{\url}[1]{#1}
\csname url@samestyle\endcsname
\providecommand{\newblock}{\relax}
\providecommand{\bibinfo}[2]{#2}
\providecommand{\BIBentrySTDinterwordspacing}{\spaceskip=0pt\relax}
\providecommand{\BIBentryALTinterwordstretchfactor}{4}
\providecommand{\BIBentryALTinterwordspacing}{\spaceskip=\fontdimen2\font plus
\BIBentryALTinterwordstretchfactor\fontdimen3\font minus
  \fontdimen4\font\relax}
\providecommand{\BIBforeignlanguage}[2]{{%
\expandafter\ifx\csname l@#1\endcsname\relax
\typeout{** WARNING: IEEEtran.bst: No hyphenation pattern has been}%
\typeout{** loaded for the language `#1'. Using the pattern for}%
\typeout{** the default language instead.}%
\else
\language=\csname l@#1\endcsname
\fi
#2}}
\providecommand{\BIBdecl}{\relax}
\BIBdecl

\bibitem{eu_ai_act_2024a}
{The European Parliament and the Council of the European Union}, ``{The AI Act:
  Regulation (EU) 2024/1689 laying down harmonised rules on artificial
  intelligence},'' 2024,
  \url{https://digital-strategy.ec.europa.eu/en/policies/regulatory-framework-ai}.

\bibitem{TOWNSEND2025PGYNC}
B.~Townsend, K.~J. Parnell, S.~G. Yaman, G.~Nemirovsky, and R.~Calinescu,
  ``{Normative conflict resolution through human–autonomous agent
  interaction},'' \emph{Journal of Responsible Technology}, vol.~21, p. 100114,
  2025.

\bibitem{troquard2024social}
N.~Troquard, M.~De~Sanctis, P.~Inverardi, P.~Pelliccione, and G.~L. Scoccia,
  ``{Social, Legal, Ethical, Empathetic, and Cultural Rules: Compilation and
  Reasoning},'' \emph{AAAI Conference on Artificial Intelligence}, vol.~38,
  no.~20, pp. 22\,385--22\,392, 2024.

\bibitem{Getir-Yaman-et-al-23}
S.~Getir-Yaman, C.~Burholt, M.~Jones, R.~Calinescu, and A.~Cavalcanti,
  ``{Specification and Validation of Normative Rules for Autonomous Agents},''
  in \emph{Proc. of {FASE}'2023.}, ser. LNCS, 2023.

\bibitem{GETIRYAMAN2025RCCPT}
S.~{Getir Yaman}, P.~Ribeiro, A.~Cavalcanti, R.~Calinescu, C.~Paterson, and
  B.~Townsend, ``Specification, validation and verification of social, legal,
  ethical, empathetic and cultural requirements for autonomous agents,''
  \emph{Journal of Systems and Software}, vol. 220, p. 112229, 2025.

\bibitem{feng-et-al-24-e}
\BIBentryALTinterwordspacing
N.~Feng, L.~Marsso, S.~Getir-Yaman, B.~Townsend, Y.~Baatartogtokh, R.~Ayad,
  V.~O. de~Mello, I.~Standen, I.~Stefanakos, C.~Imrie, G.~Rodrigues,
  A.~Cavalcanti, R.~Calinescu, and M.~Chechik, ``{Analyzing and Debugging
  Normative Requirements via Satisfiability Checking},'' in \emph{Proceedings
  of {ICSE'2024}.}\hskip 1em plus 0.5em minus 0.4em\relax ACM, 2024. [Online].
  Available: \url{https://doi.org/10.48550/arXiv.2401.05673}
\BIBentrySTDinterwordspacing

\bibitem{feng-et-al-24bb}
N.~Feng, L.~Marsso, and et. al, ``{Normative Requirements Operationalization
  with Large Language Models},'' in \emph{Proc. of {RE'2024}.}\hskip 1em plus
  0.5em minus 0.4em\relax IEEE, 2024.

\bibitem{FDR}
\url{https://cocotec.io/fdr/}.

\bibitem{SLEEC-LLM}
\url{https://github.com/alexkleijwegt/SLEEC-LLM}.

\end{thebibliography}

\appendices
\section{Description of the Demonstration}
\label{appendix:demo}
We present the demonstration in three parts, with a video available online\footnote{\url{https://www.youtube.com/watch?v=t-HBrBgVB4o}}. In the first part, we introduce the ALMI robot case study, provide a brief background on the SLEEC grammar, and present the SLEEC requirements for ALMI written in the SLEEC-DSL. These requirements are analyzed using SLEECVAL, and the results are presented. In the second part, we discuss the challenges SLEEC experts face when resolving conflicts and redundancies. To address these, we explain the intuition behind our tool, focusing on the context and templates required for the LLM. Finally, we demonstrate the tool’s concrete output, discuss threats to the validity of using LLMs, and present results from our user study.

\subsection{ALMI Robot in SLEEC DSL}
\begin{figure}
\centering
\fbox{\includegraphics[width=1\columnwidth]{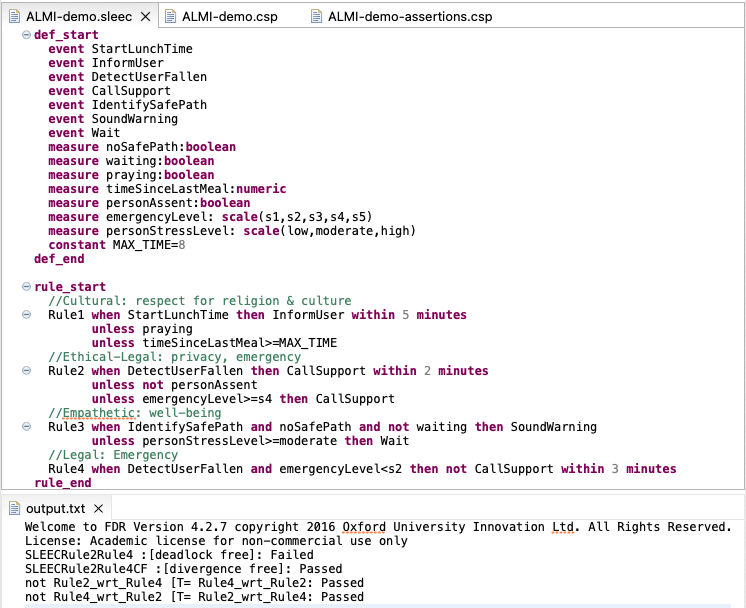}}
\caption{SLEEC Specification for ALMI robot in SLEEC DSL using SLEECVAL}\label{fig:almi-ss}
\end{figure}

In this part of the demonstration, we will introduce an agent from robotic assistive care domain called ALMI. ALMI robot is a home robot that is designed to assist elderly people for daily tasks e.g. cooking, medication taking. We will present the specification of its requirements in \sleec\ as shown in Fig.~\ref{fig:almi-ss}, explaining the intuition behind the rules. 
We will explain the syntax of the SLEEC DSL and the language elements 
using the ALMI specification which includes The definition block (surrounded by the keyword pair \sleeckeyword{def\_start, def\_end}). There are seven \sleeckeyword{event}s, seven \sleeckeyword{measure}s and one \sleeckeyword{numeric constant} defined.  The SLEEC \textbf{rules} within the \textbf{ruleBlock} will be explained together with their 
elements---\textbf{triggers}, \textbf{responses}, and \textbf{defeaters}.




%


We then present the results including counterexamples of the requirements analysis produced by the FDR4 tool, 
which illustrate a deadlock caused by the conflicting rules, Rule 2 and Rule 4.


\subsection{LLM architecture and tool output}
\begin{figure}
    \centering
\includegraphics[width=0.51\textwidth]{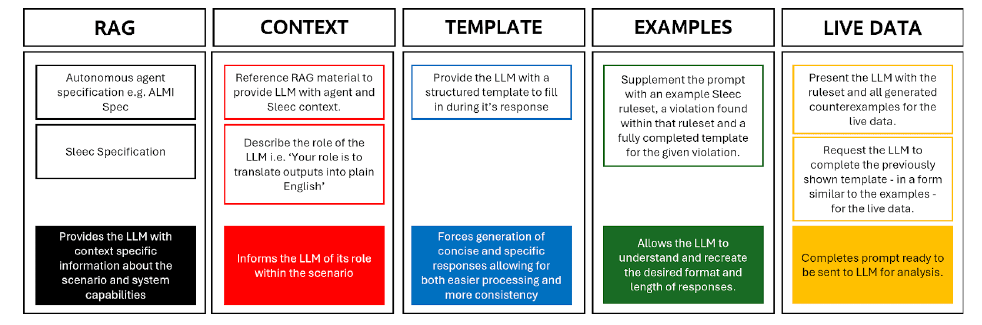}
    \caption{\small{Prompt architecture}}
    \label{fig:arch}
\end{figure}

The SLEEC-LLM tool demonstration will begin by presenting its high-level architecture, highlighting its integration with SLEECVAL and the options available to a user within the graphical interface. We will explain how counterexamples produced by the SLEECVAL analysis are passed to the LLM via a specifically designed prompt, that includes the formal output trace, the relevant SLEEC definitions and ruleset, and supplemental natural language context (Figure~\ref{fig:arch}).
Next we will walk through the exact structure of the prompt, including the JSON-style output templates used to organise and improve the LLM's response as shown in Listing \ref{lst:output}.
We will then present a live example using the ALMI case study, in which we will submit a ruleset containing two purposely contradicting rules about the systems response to an emergency situation. We will demonstrate how the SLEEC-LLM tool interprets the generated counterexample and returns to the user: An explanation of the scenario in natural language, the specific rules involved, a categorisation of the conflict type as well as two possible solutions to resolve the issue.
Lastly we will demonstrate how the outputs of the SLEEC-LLM tool can be analysed by non-technical stakeholders, including using the error information to understand the causes of the rule conflict and utilising the resolution suggestions to solve the conflict within the SLEEC ruleset.
Throughout the demonstration we will emphasise how the tool makes the SLEECVAL outputs accessible to non-technical stakeholders and also supports them in making informed decisions to resolve the errors.

\begin{figure*}[t]
    \centering
    \includegraphics[width=0.7\textwidth]{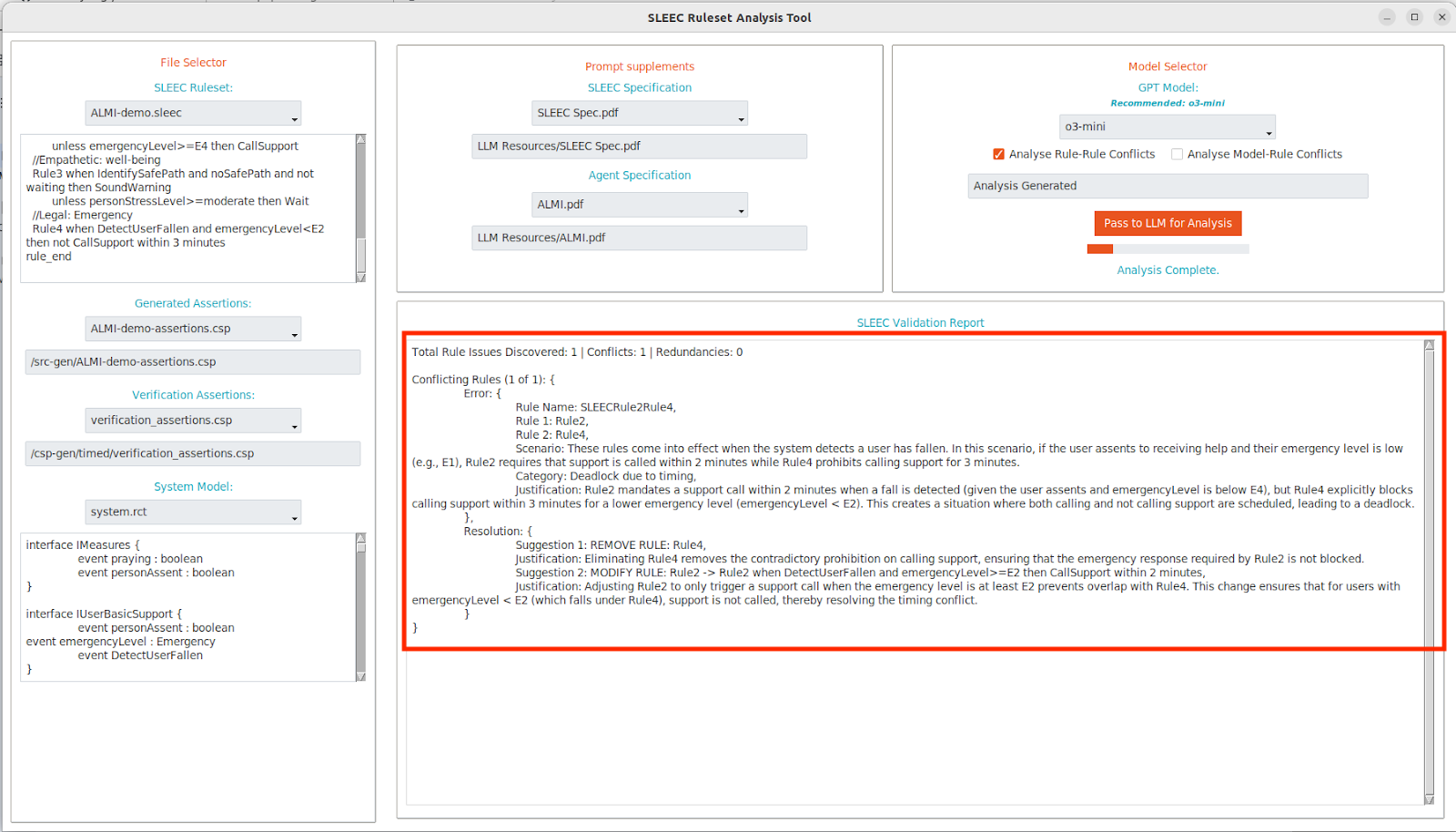}
    \caption{Screenshot of the SLEEC-LLM tool showing conflicts between Rule2 and Rule4 in the SLEEC DSL shown in~\ref{fig:almi-ss}.}
    \label{fig:ss3}
\end{figure*}


\subsection{Preliminary user study}
We will present a summary of a preliminary user study conducted to evaluate the effectiveness of the SLEEC-LLM tool and discuss with the audience. They study consisted of two small, multidisciplinary groups of participants (e.g. ethicists, laywers), each tasked with performing rule-debugging tasks on a SLEEC ruleset - once using the current SLEECVAL toolset, and once with the assistance of the LLM tool.

\begin{figure}
    \centering
    \includegraphics[width=0.45\textwidth]{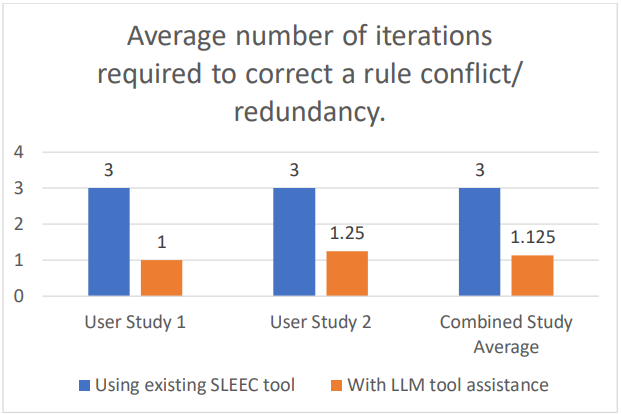}
    \caption{Average number of iterations required to correct a conflict or redundancy.}
    \label{fig:iter}
\end{figure}

\begin{figure}
    \centering
    \includegraphics[width=0.45\textwidth]{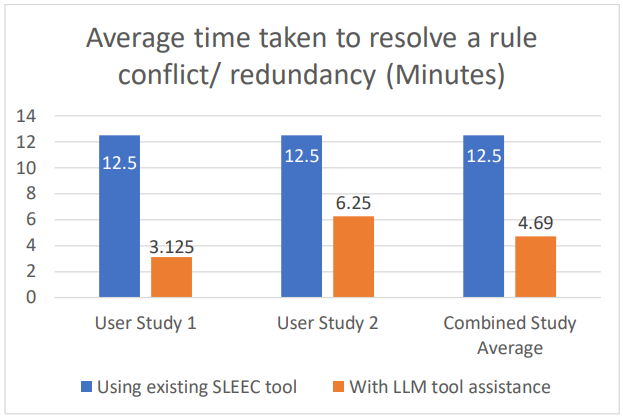}
    \caption{Average time taken to correct a conflict or redundancy.}
    \label{fig:time}
\end{figure}

During the demonstration, we will highlight the key findings from the study, focusing on the improvements in conflict resolution when the LLM tool was used. In both studies, participants using the LLM tool successfully resolved all rule conflicts and redundancies within the given 25-minute timeframe. In contrast, when fixing errors manually, participants were only able to resolve half of the ruleset errors. A 63\% reduction in the average time taken to fix an error, as well as a 63\% reduction in number of iterations required were noted during the user study. These findings are demonstrated by the graphs in Fig.~\ref{fig:iter} and Fig.~\ref{fig:time}.

We will also briefly discuss the feedback gathered from the participants, which highlighted increased confidence when using the LLM tool. The feedback from the participants were positive overall and they found the explanations and suggestions helpful, but experienced some issues with syntax correctness. This feedback will guide future improvements in the prompt design and increase the usefulness of the tool.

\subsection{Conclusion}
We will conclude indicating future work and possible extensions. We will discuss the threats to validity that may come from LLM suggestions to the users. 

\end{document}